\documentclass{article}

\font\sqi=cmssq8
\def\DR{\rm I\kern-1.45pt\rm R}
\def\DC{\kern2pt {\hbox{\sqi I}}\kern-4.2pt\rm C}
\newcommand{\ben}{\begin{enumerate}}
\newcommand{\een}{\end{enumerate}}
\newcommand{\beq}{\begin{equation}}
\newcommand{\eeq}{\end{equation}}
\newcommand{\bea}{\begin{eqnarray}}
\newcommand{\eea}{\end{eqnarray}}

\def\DH{\rm I\kern-1.5pt\rm H\kern-1.5pt\rm I}

\def\bbbz{{\mathchoice {\hbox{$\sf\textstyle Z\kern-0.4em Z$}}
{\hbox{$\sf\textstyle Z\kern-0.4em Z$}} {\hbox{$\sf\scriptstyle
Z\kern-0.3em Z$}} {\hbox{$\sf\scriptscriptstyle Z\kern-0.2em
Z$}}}}

\def\bbbz{{\rm I\!Z}} 
\chardef\ii="10
\begin{document}
\begin{center}
{\large\bf Oscillator potential  for the  four-dimensional Hall
effect
} \\
\vspace{0.4 cm} {\large Levon Mardoyan and Armen
Nersessian }\\
\vspace{0.2 cm}
{\it  Artsakh State
University, Stepanakert} \&
{\it Yerevan State University, Yerevan, Armenia}\end{center}
\begin{abstract}
We suggest the exactly solvable model of oscillator
 on the four-dimensional sphere interacting with the
 $SU(2)$ Yang monopole. We show, that the properties
of the model essentially depend on the monopole charge.
\end{abstract}
\begin{center}
\end{center}
\setcounter{equation}0
\subsubsection*{Introduction}
 The isotropic oscillator  on the  $N$-dimensional Eulclidean space
 $\DR^N$ is the distinguished system due to enormous number of symmetries.
Besides the
rotational symmetry  $so(N)$,  the oscillator  possesses  also hidden
ones, so that the
whole symmetry algebra is $su(N)$. A huge
number  of symmetries allows to construct the
generalizations of the oscillator on the curved spaces, which
inherit many properties
 of the initial system.
The   oscillator on the sphere was suggested
by Higgs \cite{higgs}. It  is defined by the potential
(through the text we assume $\mu=r_0=\hbar=1$, where $\mu$
is the mass of the particle, $r_0$ is the radius of the sphere and/or
complex end quaternionic projectyive spaces,
and $\hbar$ is Planck constant)
\beq
 V_{S^N}=\frac{\omega^2}{2}\frac{x_a x_a}{x_0^2}\label{higgs}\eeq
  where $x_a, x_0$
are the Euclidean coordinates of the ambient space $\DR^{N+1}$: $
x^2_0+x_a x_a=1$.
This system inherits the rotational symmetries of the flat
oscillator and the  hidden
symmetries as well. However,
in contrast to the  oscillator on $\DR^N$, the whole symmetry algebra
of the spherical (Higgs) oscillator is nonlinear one.
The oscillator on  the complex projective space
 $\DC P^N$ is  defined by the potential \cite{cpn}
\beq
V_{\DC P^N}=\omega^2  z_a\bar z_a=
\alpha^2\frac{u_a\bar u_a}{u_0\bar u_0},\quad
u_a\bar u_a+u_0\bar u_0=1,
\label{bn}\eeq
where  $z_a=u_a/u_0$ are
inhomogeneous coordinates of $\DC P^N$.
For  $N=1$, i.e. in the case of two-dimensional sphere $\DC
P^1 =S^2$, the  suggested system has no hidden symmetries, as opposed
with the Higgs oscillator on $S^2$. However,
after inclusion of constant magnetic field,
the $\DC P^1$-oscillator remains exactly solvable, while
 the Higgs oscillator  looses its hidden symmetries and
 exact solvability property as well.
 Moreover, for  $N>1$ the $\DC P^N$ oscillator has a hidden symmetries
  preserving after inclusion of the constant
magnetic field \cite{qcpn}.
Looking on (\ref{higgs}) and (\ref{bn}), one can observe, that
 the expression of the $S^N$-oscillator potential
in terms of the ambient space   $\DR^{N+1}$ is  very similar to
the one for the $\DC P^N$-oscillator potential in terms of the ``ambient"
space $\DC^{N+1}$.
Continuing this sequence, we   define  the oscillator potential
on the quaternionic projective spaces $\DH P^N$
as follows
\beq
V_{\DH P^N}=\omega^2 w_a{\bar w}_a\;, \quad
w_a=\frac{{\bf u}_a}{{\bf u_0}}=\frac{u_{a|1}+ ju_{a|2}}{u_{0|1}+ju_{0|2}}.
\label{hpn}\eeq
 Here $w_a$ are inhomogeneous (quaternionic) coordinates of the
quaternionic projective space $\DH P^N$, and ${\bf u}_a=u_{a|0}+ju_{a|1},
{\bf u}_0=u_{0|1}+ju_{0|2}$ are  the Euclidean coordinates of ``ambient"
quaternionic space $\DH^{N+1}=\DC^{2N+2}$,
${\bf u}_a{\bf\bar u}_a +{\bf u}_0{\bf \bar u}_0=1$.
In analogy with $\DC P^N$-oscillator one can  expect,
 that this system could be exactly solvable also in the
 presence of $SU(2)$ instanton field.
 In this note we shall show, that  for  $\DH P^1=S^4$
it is indeed the case.

Precisely, {\sl we  propose the exactly solvable model of oscillator on
four-dimensional sphere $S^4$, interacting with SU(2) Yang monopole located
in the center of $S^4$.}

This model  seems to be applicable in the
theory of  four-dimensional Hall effect suggested
 by Zhang and Hu \cite{4h}.
Zhang-Hu theory is based on
the quantum mechanics of colored particle moving on a
four-dimensional sphere in the field of $SU(2)$ Yang monopole \cite{yang},
 and has effective  three-dimensional edge dynamics.
Let us remind, that the theory of  conventional (two-dimensional)
 Hall effect
is related with the first Hopf map $S^3/S^1=S^2\cong\DC P^1$,
while the Zhang-Hu theory  is related with the second Hopf map
$S^7/S^3= S^4\cong\DH P^1$.
In fact, it could be viewed as a ``quaternionic'' analog of
conventional quantum Hall effect\footnote{The ``octonionic"   Hall
effect  based in the third  Hopf map
$S^{15}/S^7=S^8$  was suggested in Ref. \cite{octonions},
while  and the  the Hall effects on $\DC P^N$ and $S^3$
have been considered in \cite{kn} and \cite{kns3}, respectively.
The discussion of other  aspects
of higher-dimensional Hall effect
 could be found in Refs. \cite{4other}.}.
 In the Zhang-Hu model  key role plays the  field of Yang monopole
 preserving the $SO(5)$ symmetry of the system and providing it
 with degenerate ground state.
While transition to the effective three-dimensional edge theory assumes
the introduction of confining potential breaking the $SO(5)$ symmetry
 of the system.
In proposed  oscillator model, in spite of the presence of confining
(oscillator) potential breaking the $SO(5)$ symmetry of the system,
the ground state remains degenerated,
and the system is still  exactly solvable.
Hence, one can develop the suitable
modification of the four-dimensional Hall
effect, where the confining potential appears  {\sl at hoc}.
Notice, that the Hall effect on $\DR^4$ with the oscillator potential
has been considered
by Elvang and Polchinski \cite{polchinski}.
Proposed model gives the opportunity to consider similar theory on
$S^4$ and  containing Elvang-Polchinski model as a limiting case.

\subsubsection*{The model}
Let us going to formulate the model. Due to the second Hopf map
the Euclidean coordinates of $\DR^5$ and $\DH^2=\DC^4$
are related as follows
\beq
w (\equiv x_1+ix_2+jx_3+kx_4)=2{\bf u}_1{\bf \bar u}_2, \
; x_0={\bf u}_1{\bf \bar u}_1-{\bf u}_2{\bf \bar u}_2,
\quad \Rightarrow\quad x^2_0+w\bar w=
({\bf u}_1{\bf \bar u}_1+{\bf u}_2{\bf \bar u}_2)^2 \;(\equiv r^2)\;.
\eeq
Also, we introduce  the ``hyperspherical" coordinates
\beq
x_0 = r\cos \theta ,\;
x_1 + ix_2 = r \sin \theta \cos \frac{\beta}{2}{\rm e}^{i\frac{\alpha
+\gamma}{2}}, \;
x_3 + ix_4 = r \sin \theta \sin \frac{\beta}{2}{\rm e}^{i\frac{\alpha -
\gamma}{2}},\quad
\theta,\beta \in [0,\pi],\; \alpha \in [0,2\pi),
\;\gamma \in [0,4\pi).
\eeq
In these terms the  potential (\ref{pot+}) reads
\beq
V^{+}_{S^4}=\frac{\omega^2}{2}\frac{r-x_0}{r+x_0}=
\frac{\omega^2}{2}\frac{1-\cos\theta}{1+\cos\theta}.
\label{pot+}\eeq
The coupling  to the monopole field is performed in a minimal
way, $-i\partial/\partial x^A\to
-i\partial/\partial x^A+A_A^aT_a $,
where $T_a$ are the $SU(2)$ generators on the internal space $S^2$
of Yang monopole, $[\hat T_a,\hat T_b] = i\epsilon_{abc}\hat T_c $,
and $A^a_A$ is the connection defining Yang monopole.
In these terms the quantum Hamiltonian
looks as follows
 \footnote{Here and further we omit the details of calculations,
 referring to the papers \cite{mardoyan}
 devoted to the study of five-dimensional Coulomb problem in
 the presence of $SU(2)$ Yang monopole. The kinetic part of our Hamiltonian
 is simply spherical part of that system.  calculations
 of wavefunctions and spectrum of our system are
 very closed to those performed in the mentioned papers.}
\beq
{\cal H}=\frac{1}{2}\left[
\frac{1}{ \sin^3 \theta}\frac{\partial}{\partial \theta}
\left(\sin^3 \theta \frac{\partial}{\partial \theta}\right) -
\frac{{\hat L}^2}{\sin^2 \theta/2} -
\frac{{\hat J}^2}{\cos^2 \theta/2}\right]+
\frac{\omega^2}{2}\frac{1-\cos\theta}{1+\cos\theta}.
\eeq
Here ${\hat L}_a$ are the components of $SU(2)$ momentum
$[\hat L_a,\hat L_b] = i\epsilon_{abc}\hat L_c $,
\beq
{\hat L}_1 = i\left(\cos{\alpha}\cot{\beta}
\frac{\partial}{\partial \alpha}+
\sin{\alpha}\frac{\partial}{\partial \beta} -
\frac{\cos{\alpha}}{\sin{\beta}}
\frac{\partial}{\partial \gamma}\right), \;
{\hat L}_2 = i\left(\sin{\alpha}\cot{\beta}
\frac{\partial}{\partial \alpha} -
\cos{\alpha}\frac{\partial}{\partial \beta} -
\frac{\sin{\alpha}}{\sin{\beta}}
\frac{\partial}{\partial \gamma}\right) ,\;
{\hat L}_3 = -i\frac{\partial}{\partial \alpha}.
\label{L}\eeq
and
$\hat J_a = \hat L_a + \hat T_a$,
\beq
[\hat L_a,\hat L_b] = i\epsilon_{abc}\hat L_c,\,\,\,\,\,\,
[\hat L_a,\hat J_b] = i\epsilon_{abc}\hat L_c,\,\,\,\,\,\,
[\hat J_a,\hat J_b] = i\epsilon_{abc}\hat J_c,
\eeq
It is convenient to represent the generators $T^a$
in terms of $S^3$, as in  (\ref{L})(where instead of $\alpha,\beta,\gamma$
appear the coordinates of $S^3$, $\alpha_T,\beta_T,\gamma_T$), with the
following condition imposed
\beq
{\hat T}^2\Psi(\alpha,\beta,\gamma,\theta,\alpha_T,\beta_T,\gamma_T)=
T(T+1)\Psi (\alpha,\beta,\gamma,\theta,\alpha_T,\beta_T,\gamma_T).\eeq
This conditions corresponds to the fixation of the isospin $T$,
 and restricts
the configuration space of the system from
$S^4\times S^3=S^7$ to
$S^4\times S^2=\DC P^3$.
Notice, that the  generators $\hat J_a$,  $\hat L^2$, $\hat T^2$
are constants of motion,
while $\hat L_a$, $\hat T_a$ do not commute with the Hamiltonian.

To  solve the  Schr\"odinger equation
${\cal H}\Psi={\cal E}\Psi$, we introduce the separation ansatz
\beq
\Psi(\theta,\alpha,\beta,\gamma, \alpha_T,\beta_T,\gamma_T) =
Z(\theta )\Phi(\alpha,\beta,\gamma,\alpha_T,\alpha_T, \gamma_T).
\label{sep}\eeq
where $\Phi$ are the eigenfunctions of ${\hat L}^2$, ${\hat T}^2$ and
${\hat J}^2$ with the eigenvalues $L(L+1)$, $T(T+1)$ and $J(J+1)$.
So, $\Phi$ could be represented in the form
\beq
\Phi =
\sum_{M=m+t}\left(JM|L,m';T,t'\right)D_{mm'}^L(\alpha,\beta,\gamma)
D_{tt'}^T(\alpha_T,\beta_T,\gamma_T)
\eeq
where $\left(JM|L,m';T,t'\right)$ are the Clebsh-Gordan coefficients
and $D_{mm'}^L$ and $D_{tt'}^T$ are the Wigner functions.

Using  the above separation anzats, we get
 the following``radial" Schr\"odinger equation
\begin{eqnarray}
\frac{1}{\sin^3\theta}\frac{d}{d\theta}\left(\sin^3\theta
\frac{dZ}{d\theta}\right)-\frac{2L(L+1)}{1-\cos\theta}Z-
\frac{2J(J+1)}{1+\cos\theta}Z+
2\left(
{\cal E}-\frac{\omega^2}{2}\frac{1-\cos\theta}{1+\cos\theta}\right)Z=0,
\label{2.4.1}
\end{eqnarray}
%
It is convenient to transit to the new variable $y=(1-\cos
\theta)/2$, $y\in[0,1)$ and to represent the radial wavefunction
in the following form
\beq
Z(y) = y^L(1-y)^{\widetilde J}W(y)\quad {\rm
where}\quad
{\widetilde J}({\widetilde J}+1)\equiv J(J+1)+\omega^2
\eeq
In this terms the radial Schr\"odinger equation looks as follows
\beq
y(y-1)\frac{d^2W}{dy^2} +
 \left[2L+2-(2{\widetilde J}+2L+4)y\right]\frac{dW}{dy}-
\left[({\widetilde J}+L)({\widetilde J}+L+3)-2{\cal E} -\omega^2\right]W=0.
\eeq
The regular solution of this equation is confluent hypergeometric
function $W={_2}F_1(-n,n +2L+2{\widetilde J}+3,2L+2,y)$, where
 $-n=L+{\widetilde J}+3/2 -\sqrt{2{\cal E}+\omega^2+9/4}$.
Hence, the energy spectrum reads
\beq
{\cal E}=
\frac{1}{2}(n+L+{\widetilde J})(n+L+{\widetilde J}+3)-
\frac{\omega^2}{2},\qquad
n=0,1,2,\ldots
 \eeq
 and the regular wavefunction is defined by the expression
\beq
Z(\theta)= 
\sqrt{\frac
{(2n+2L+2{\widetilde J}+3)n!\Gamma(n+2L+2{\widetilde J}+3)}
{2^{2L+2{\widetilde J}+3}\Gamma(n+2L+2)\Gamma(n +2{\widetilde J}+2)}}
(1-\cos \theta)^L(1+\cos
\theta)^{\widetilde J} P_{n}^{(2L+1,2{\widetilde J}+1)}(\cos \theta).
\label{wf}\eeq
Let us remind, that
\beq
J=|L-T|,|L-T|+1,...,L+T, \qquad
L=0, {1}/{2}, 1,\ldots
\label{LJdef}\eeq
In the absence of the potential, $\alpha^2=0$, one has ${\widetilde J}=J$.
In this case one can introduce the principal quantum
number ${\cal N}=n+J+L$, and recover standard expressions for the
the spectrum and wavefunctions
of the free particle on $S^4$ moving in the field of Yang monopole.
Also, in this case the Hamiltonian is invariant under the change
 \beq
\theta\to \pi -\theta,\quad L\to J, \quad J\to L.
\label{ma0}\eeq
In the absence of  monopole, $T=0$, one has
$J=L$, and the model is symmetric under above transformation
in the presence of oscillator potential as well.
\\
However, when  the monopole and potential fields
 simultaneously appear, the situation is
 essentially different.
The radial Schr\"odiger equation (\ref{2.4.1}) becomes invariant
under the changes
 \beq
\theta\to \pi-\theta,\quad L\to J,
\quad {\widetilde J}\to {\widetilde  L},
\quad {\rm where}
\quad{\widetilde L}({\widetilde L}+1)\equiv L(L+1)+\omega^2.
\label{ma1}\eeq
Hence, the particle on $S^4$ moving in the fields of  Yang
 monopole and of the potential
\beq
V^{-}_{S^4}=
\frac{\omega^2}{2}\frac{1+\cos\theta}{1-\cos\theta},
\label{pot-}\eeq
has the following spectrum
\beq
{\cal E}=
\frac{1}{2}(n+J+{\widetilde L})(n+J+{\widetilde L}+3)-
\frac{\omega^2}{2},\quad
{\widetilde L}({\widetilde L}+1)\equiv L(L+1)+\omega^2\; .
\eeq
Its wavefunction could be also found from (\ref{wf}) by the change
 (\ref{ma1}).
%
In this case, the planar limit should be taken in the vicinity of the
{\sl north pole} of the four-dimensional sphere, in contrast with
 previous consideration.

One can combine the potentials (\ref{pot+})
and (\ref{pot-}) in the following one
\beq
V^{s}_{S^4}=
\frac{\omega^2_1}{2}\frac{1+\cos\theta}{1-\cos\theta}+
\frac{\omega^2_2}{2}\frac{1-\cos\theta}{1+\cos\theta}.
\label{pots}\eeq
In the presence of monopole field
 the system with this potential has a spectrum
\beq
{\cal E}^{s}=
\frac{1}{2}(n+{\widetilde J}+{\widetilde L})
(n+{\widetilde J}+{\widetilde L}+3)-
\frac{\omega^2_1+\omega^2_2}{2},\quad
{\widetilde L}({\widetilde L}+1)\equiv L(L+1)+\omega^2_1,\quad
{\widetilde J}({\widetilde J}+1)\equiv J(J+1)+\omega^2_2
\qquad\quad n=0,1,2,\ldots,
\eeq
When $\omega_1=\omega_2$, the system is
invariant under reflection $\theta-\pi-\theta$.
But the prise is singularity of the potential in both poles.
In the flat limit this system results in the singular oscillator on $S^4$.

Now, taking into account, that transformation of the quaternionic
coordinate $w\to 1/w$ (and, consequently, the reflection
$\theta\to\pi-\theta$) corresponds to the transition from
the monopole to anti-monopole (i.e.  from the $SU(2)$ monopole with
topological charge $+1$ to the one with topological charge $-1$)
(see, e.g., \cite{atiah}), we conclude,
that the spectrum of the proposed oscillator model
essentially depends on the topological charge of monopole.
The same  phenomenon  on $\DR^4$ has been
observed in Ref. \cite{polchinski}.
\subsubsection*{Discussion}
We constructed the exactly solvable oscillator
on four-dimensional sphere interacting with Yang monopole.
The presence of monopole provides  the system with the degenerate
 ground state, which allows to wish that the system could be useful in the
 four-dimensional Hall effect.
 An interesting peculiarity of the system is the essential
 dependence of its spectrum from the topological charge of Yang monopole.
This asymmetry (with respect to monopole and antimonopole)
looks  very similar to the  behaviour of the noncommutative
quantum mechanics on $\DR^2$ and $S^2$ in the  constant
magnetic field \cite{nc}. It seems, that one can established the explicit
 correspondence  between these  two pictures.
It is clear, that observed asymmetry is not a specific property of
the oscillator potential, suggested in this note, but of
 any potential $V(\theta )$, which is not invariant under reflections
$\theta \to \pi-\theta $.

Let us also mention, that the expession  of the
$\DC P^1$-oscillator expresses in terms of $\DR^3$is identical
to the expression of the suggested $\DH P^1$-oscillator
potential in terms of the ambient
$\DR^5$ space. Clearly, this is due to the isomorphisms $\DC P^1= S^2$
and $\DH P^1 =S^4$ and the Hopf maps. So, the same potential on the
$S^8$ would define the oscillator model respecting the interaction with
$SO(8)$ monopole field.

\end{document}